\documentclass[review]{elsarticle}

\usepackage{hyperref}

\journal{Journal of \LaTeX\ Templates}

\begin{document}

\begin{frontmatter}

\title{ALPHA AND CLUSTER DECAY HALF LIVES IN TUNGSTEN ISOTOPES: A MICROSCOPIC ANALYSIS}

\author{NITHU ASHOK\corref{nit}}
\cortext[nit]{Corresponding author}
\ead[e-mail]{nithu.ashok@gmail.com}
\author{ANTONY JOSEPH}
\address{Department of Physics, University of Calicut,\\
 Kerala, India}




\begin{abstract}
Alpha and cluster decay half-lives for Tungsten (W) isotopes in the range between 2p drip line and beta stability line are studied.  The sensitivity of different Skyrme parametrizations in predicting the alpha-decay and probable cluster decay modes from W isotopes have been analysed.  The half-lives are calculated using UDL.  Predicted half-lives are compared with ELDM and also with available experimental values.  The use of HO and THO basis do not produce much differences in the results. The study also revealed the role of neutron shell closure in cluster decay process.
\end{abstract}

\begin{keyword}
\texttt{cluster; Q-value; half-life, Hartree-Fock-Bogoliubov.}
\end{keyword}

\end{frontmatter}


\section{Introduction}

Cluster radioactivity is defined as the spontaneous emission of a fragment, heavier than alpha particle and lighter than the lightest fission fragment, from the parent nuclei, without being accompanied by neutron emission.  This phenomenon  was first predicted by Sandulescu et.al in 1980\cite{sand}.  This exotic decay was later experimentally observed by Rose and Jones\cite{rose} in 1984, with the emission of $^{14}\textrm{C}$ cluster from $^{223}\textrm{Ra}$.  Cluster radioactivity is a cold nuclear phenomenon, explained based on quantum mechanical fragmentation theory (QMFT)\cite{raj,qmft}.  One of the dominant decay modes in nuclei is the $\alpha$-decay, which is the emission of $_{2}^{4}\textrm{He}$ from the parent nuclei.  The probability of formation of a cluster is mainly determined by its binding energy.  This implies that of all the possible cluster emissions, $\alpha$-cluster is the most prominent one.  

Many theoretical models have been developed to study the phenomenon of cluster radioactivity. The widely used phemenological models are Preformed Cluster Model (PCM)\cite{pcm} and Unified Fission Model (UFM)\cite{ufm}.  In PCM, the cluster is assumed to be preformed inside the parent nucleus and the preformation probability has to be found out explicitly.  In UFM, the parent nucleus undergoes continuous dynamical changes through a molecular phase and finally disintegrates into a daughter and a cluster.  Here the preformation probability is taken as unity.  Several theoretical and experimental studies on cluster decay have been carried out in recent years.  Different studies show that this phenomenon occurs in those regions where daughter nuclei should either be doubly magic or in its vicinity.

In our previous work, we have analysed the different decay modes in Os isotopes using Hartree-Fock-Bogoliubov (HFB) theory\cite{os}.  In the present work, the feasibility of alpha and cluster decay from W isotopes have been studied using HFB theory.  Many works, both theoretical and experimental, have been devoted to the study of alpha decay from various W nuclei in recent years\cite{rev1,rev2,rev3}.  Here we made an attempt to study the different cluster decay modes from W isotopes in a systematic way with the help of Skyrme HFB theory.

The paper is organised as follows.  In sec. 2, a brief account of the microscopic theory (HFB theory) which is used for the present study is given.  In sec. 3 we have shown the details of our calculations.  Results and discussion are given in sec. 4, where we have presented the main part of the study. In sec. 5, the conclusion drawn from the present work is given.  
\section{Hartree-Fock-Bogoliubov Theory}
A brief description of the Hartree-Fock-Bogoliubov theory is given below.  
The many body Hamiltonian expressed in terms of annhilation and creation operators is given by\cite{ring},
\begin{equation}
 H=\Sigma _{ij} t_{ij} a_{i}^{\dagger}a_{j}+ \frac{1}{4}\Sigma_{ijkl} V_{ijkl} a_{i}^{\dagger}a_{j}^{\dagger}a_{k}a_{l}
\end{equation}
A set of quasiparticle state is used as the trial wave function.  The bare particles are transformed to quasiparticles by using Bogoliubov transformation\cite{ring}:
\begin{equation}
  \beta_{k}^{\dagger}=\Sigma _{l}U_{lk}a_{l}^{\dagger}+V_{lk}a_{l}
\end{equation} 
\begin{equation}
\beta_{k}=\Sigma _{l}V_{lk}^{*}a_{l}+U_{lk}^{*}a_{l}^{\dagger}
\end{equation}
In terms of the density matrix $\rho $ and the pairing tensor $\kappa $, on which the wavefunction $\Phi $ depends, the Hartree-Fock-Bogoliubov energy can be expressed as
\begin{equation}
E[\rho ,\kappa ]=\frac{<\Phi |H-\lambda N|\Phi >}{<\Phi |\Phi >}=Tr[(\varepsilon +\frac{1}{2}\Gamma )\rho ]-\frac{1}{2}Tr[\Delta \kappa ^{*}]
\end{equation}
where Hartree Fock(HF) potential $\Gamma $ and pairing potential $\Delta $ are defined as
\begin{equation}
\Gamma_{kl} = \Sigma _{i,j}\bar{v}_{kjli}\rho _{ij}
\end{equation}
\begin{equation}
\Delta _{kl} =\frac{1}{2}\Sigma _{i,j}\bar{v}_{kjli}\kappa  _{ij}
\end{equation}
The HFB equations are obtained by applying the variational principle to $(H-\lambda N)$.  In the matrix form, the HFB equation is given by\cite{bend},
    \begin{equation}
     \left( {\begin{array}{cc}
h-\lambda   &  \Delta   \\
-\Delta ^{*}      &   -h^{*}+\lambda 
\end{array} }\right)
\left( {\begin{array}{c}
U_{n}   \\
 V_{n}
\end{array} }\right)
=E_{n}\left( {\begin{array}{c}
U_{n}   \\
 V_{n}
\end{array} }\right)
     \end{equation}     
where $h=t+\Gamma $, $E_{n}$ is the quasiparticle energy and $\lambda $ is the chemical potential.

\section{Details of calculation}
Skyrme HFB equations have been solved using cylindrically deformed HO and THO basis\cite{hfbtho}.The THO set of basis wave functions consist of transformed harmonic oscillator functions, which are generated by applying the local scale transformation (LST)\cite{lst} to the HO single particle wave functions\cite{htho}.  
Numerical calculations have been carried out using 20 oscillator shells.  The cut-off energy is taken as 60 MeV.  
In the particle-hole channel, we have used the zero-range effective Skyrme interactions\cite{sky}.  In the present work we have used different Skyrme forces like SIII\cite{siii}, SkP\cite{skp}, SLy5\cite{sly5}, SkM*\cite{skm}, UNEDF0\cite{une0} and UNEDF1\cite{une1}.  These Skyrme forces are selected as they are very efficient in reproducing the ground state properties.  Also they differ in various parameters and this helps to analyse the effect caused by different factors.  They vary in the value of effective mass, surface energy, with the inclusion of $J^{2}$ term, centre of mass correction etc.  The details of the parameters can be obtained from their corresponding references.

In the particle-particle (pairing) channel, pairing interaction is included using the density dependent delta interaction\cite{dddi, dddi1} of the form\cite{mix},
\begin{equation}
  V^{n/p}_{\delta }(\vec{r_{1}},\vec{r_{2}})=V_{0}^{n/p}[1-\frac{1}{2}(\frac{\rho(\vec{r_{1}} +\vec{r_{2}})}{\rho _{0}} )^{\alpha }]\delta (\vec{r_{1}}-\vec{r_{2}})
\end{equation}
where the saturation density\cite{sat} $\rho _{0}$=0.16 fm$^{-3}$ and $\alpha $=1.

The half-lives corresponding to each decay mode is calculated using a standard formula, the Universal decay law(UDL)\cite{udl} which has been deduced from WKB approximations, with some modifications.  It is given by,  
\begin{equation}
log_{10}T_{1/2}=aZ_{c}Z_{d}\sqrt{\frac{A}{Q}}+b\sqrt{AZ_{c}Z_{d}(A_{c}^{1/3}+A_{d}^{1/3})}+c
\end{equation}
where the constants are a=0.4314, b=-0.4087 and c=-25.7725.\\ $Z_{c}$, $Z_{d}$ are the atomic number of cluster and daughter nuclei , $A_{c}$, $A_{d}$ are the mass number of cluster and daughter nuclei and 
\begin{equation}
 A=\frac{A_{c}A_{d}}{A_{c}+A_{d}}
  \end{equation}
Here Q is the Q-value of the decay.
Moreover, calculations have been limited to those decays which are having half-lives in the experimentally measurable range ie. $T_{1/2}<10^{30}$ s.
\begin{table}[ht]
\caption{\label{tab:alp}Q-values of alpha decay in even-even W isotopes calculated with Skyrme HFB equations solved using HO(top) and THO(bottom) basis along with ELDM and available experimental values.}
\resizebox{\textwidth}{!}
  {\begin{tabular}{ccccccccc} \hline
\multicolumn{1}{c}{Alpha decay} & \multicolumn{7}{c}{Q value}\\ \cline{2-9}  
& SIII & SKP & SkM* & SLy5 & UNEDF0 & UNEDF1 & ELDM  & exp  \\ \hline
$^{158}\textrm{W}\rightarrow \alpha +^{154}\textrm{Hf}$ &8.2843&7.0719&7.9058&8.3707&6.8296&7.6901&6.6051&\\ 
&8.2448&6.9272&7.9024&8.3608&6.8330&7.6648&&\\
$^{160}\textrm{W}\rightarrow \alpha +^{156}\textrm{Hf}$ &6.1670&6.1679&6.6801&6.2786&5.9269&6.1993&6.0651&\\
&6.1674&6.1565&6.6793&6.2924&5.9236&6.1915&&\\
$^{162}\textrm{W}\rightarrow \alpha +^{158}\textrm{Hf}$ &5.4312&5.5584&5.8998&5.4535&5.2839&5.3959&5.6781&$5.53^{(ref. 31)}$\\
&5.4375&5.5595&5.9009&5.4641&5.2811&5.3908&&\\
$^{164}\textrm{W}\rightarrow \alpha +^{160}\textrm{Hf}$ &5.2161&5.3343&5.5125&5.2823&5.0426&5.0211&5.2781&$5.153^{(ref. 31)}$\\
&5.2140&5.3451&5.5043&5.2808&5.0586&5.0135&&\\
$^{166}\textrm{W}\rightarrow \alpha +^{162}\textrm{Hf}$ &4.8437&5.0529&5.1203&4.8793&4.8716&4.6123&4.8561&\\
&4.8417&5.0623&5.1196&4.8811&4.8872&4.6160&&\\
$^{168}\textrm{W}\rightarrow \alpha +^{164}\textrm{Hf}$ &4.5294&4.7429&4.8816&4.5113&4.6839&4.2422&4.5001&\\
&4.5225&4.7255&4.8851&4.5205&4.6878&4.2600&&\\
$^{170}\textrm{W}\rightarrow \alpha +^{166}\textrm{Hf}$ &4.3104&4.4892&4.8444&4.2749&4.4489&3.9996&4.1441&\\
&4.3094&4.4941&4.8412&4.2828&4.4591&4.0115&&\\
$^{172}\textrm{W}\rightarrow \alpha +^{168}\textrm{Hf}$ &3.4093&4.2776&4.7594&4.0082&4.2088&-&3.8391&\\
&3.4486&4.2899&4.7566&4.0265&4.2107&-&&\\
$^{174}\textrm{W}\rightarrow \alpha +^{170}\textrm{Hf}$ &3.6332&4.0139&5.0409&3.8931&3.9625&3.6108&3.6021&\\
&3.6458&4.0131&5.0231&3.9034&3.9671&3.6265&&\\
$^{176}\textrm{W}\rightarrow \alpha +^{172}\textrm{Hf}$ &3.4270&4.0882&4.9569&4.4733&3.7381&3.6371&3.3351&\\
&3.4492&4.0904&4.9373&4.4529&3.7501&3.6459&&\\
$^{178}\textrm{W}\rightarrow \alpha +^{174}\textrm{Hf}$ &3.4460&4.0156&4.8358&4.0885&3.5356&3.6378&3.0128&\\
&3.4257&4.0211&4.8235&4.0597&3.5539&3.6392&&\\
$^{180}\textrm{W}\rightarrow \alpha +^{176}\textrm{Hf}$ &2.4381&3.6434&4.4935&3.2756&3.2196&3.2942&2.5149&$2.516^{(ref. 32)}$\\
&2.4365&3.6597&4.4771&3.2793&3.2383&3.2923&&\\ \\
\hline
\end{tabular}}
\end{table}


\begin{table*}[h]
\caption{\label{tab:sd}Comparison of standard deviation of alpha decay half-lives of W isotopes calculated for different Skyrme forces }
\centering
   {\begin{tabular}{ccccccc} \hline
& SKP & SLY5&SIII & SKM* & UNEDF0&UNEDF1\\ \hline
HO&1.0088&0.8719&0.6646&2.0293&0.9245&0.9828\\
THO&1.0278&0.8850&0.6540&2.0208&0.9281&0.9850\\
\hline
  
  \end{tabular} }
\end{table*} 


\begin{table}[!]\footnotesize
\caption{\label{tab:clus}Same as Table 1, but for various clusters}
\centering
\renewcommand{\arraystretch}{0.6}
  {\begin{tabular}{@{}*{9}{c}@{}} \hline
\multicolumn{1}{c}{Cluster decay} & \multicolumn{7}{c}{Q value}\\ \cline{2-8}  
& SIII & SKP & SkM* & SLy5 & UNEDF0 & UNEDF1 & ELDM   \\ \hline

$^{158}\textrm{W}\rightarrow ^{8}\textrm{Be} +^{150}\textrm{Yb}$&8.5165&8.6746&8.1450&9.0354&7.9249&7.5949&9.9983 \\
&8.5196&8.5584&8.1576&9.0398&7.9364&8.6686& \\
$^{160}\textrm{W}\rightarrow ^{8}\textrm{Be} +^{152}\textrm{Yb}$ &14.1145&12.7116&14.1092&14.1472&12.1651&13.3722&11.9983\\
&14.0845&12.7087&14.0999&14.1462&12.1749&13.3233& \\
$^{162}\textrm{W}\rightarrow ^{8}\textrm{Be} +^{154}\textrm{Yb}$&10.9519&11.0198&11.9117&10.8985&10.3995&10.8471&10.9903 \\
&10.9602&11.0270&11.9146&10.9209&10.3887&10.8340& \\
$^{164}\textrm{W}\rightarrow ^{8}\textrm{Be} +^{156}\textrm{Yb}$&9.8920&9.9878&10.5721&9.7219&9.3746&9.4569&10.0883 \\
&9.9013&10.0048&10.5653&9.7282&9.3851&9.4439& \\
$^{166}\textrm{W}\rightarrow ^{8}\textrm{Be} +^{158}\textrm{Yb}$&9.2564&9.4512&9.8530&9.2583&8.9077&8.6402&9.1803 \\
&9.2178&9.4752&9.8442&9.2523&8.9298&8.6371& \\
$^{168}\textrm{W}\rightarrow ^{8}\textrm{Be} +^{160}\textrm{Yb}$&8.3727&8.8660&9.2632&8.5529&8.4527&7.9136&8.3303 \\
&8.3601&8.8807&9.2667&8.5514&8.4652&7.9185& \\
$^{170}\textrm{W}\rightarrow ^{8}\textrm{Be} +^{162}\textrm{Yb}$&7.4664&8.3119&9.1163&7.9436&7.9983&7.3057&7.5953 \\
&7.4601&8.3565&9.1158&7.9457&8.0080&7.3135& \\
\\

$^{158}\textrm{W}\rightarrow ^{12}\textrm{C} +^{146}\textrm{Er}$ &18.0288&18.1709&18.3956&17.7180&17.6459&17.8516&20.622\\
&18.0469&18.0619&18.4012&17.7474&17.6648&17.8583& \\
$^{160}\textrm{W}\rightarrow ^{12}\textrm{C} +^{148}\textrm{Er}$&21.3810&21.4193&21.3479&21.8918&20.3908&21.4069&22.099 \\
&21.3880&21.4263&21.3589&21.9185&20.3914&21.3912& \\
$^{162}\textrm{W}\rightarrow ^{12}\textrm{C} +^{150}\textrm{Er}$&26.1713&24.4897&26.5842&25.8829&23.5826&25.0532&23.831 \\
&26.1485&24.4849&26.5816&25.8869&23.5932&25.0148& \\
$^{164}\textrm{W}\rightarrow ^{12}\textrm{C} +^{152}\textrm{Er}$&22.3285&22.3276&23.6660&22.1521&21.2683&21.9540&22.266 \\
&22.3298&22.3407&23.6585&22.1721&21.2761&21.9438& \\
$^{166}\textrm{W}\rightarrow ^{12}\textrm{C} +^{154}\textrm{Er}$&20.6668&20.7735&21.6400&20.4620&19.8958&19.8242&20.717 \\
&20.6676&20.7931&21.6318&20.4656&19.9143&19.8154& \\
$^{168}\textrm{W}\rightarrow ^{12}\textrm{C} +^{156}\textrm{Er}$&19.4806&19.8267&20.6571&19.6931&19.1118&18.5680&19.317 \\
&19.4665&19.8336&20.6541&19.6781&19.1288&18.5744& \\
$^{170}\textrm{W}\rightarrow ^{12}\textrm{C} +^{158}\textrm{Er}$&18.1849&19.0182&19.9845&18.7305&18.2899&17.6292&18.014 \\
&18.1726&19.0666&19.9816&18.7168&18.3092&17.649& \\
\\

$^{158}\textrm{W}\rightarrow ^{16}\textrm{O} +^{142}\textrm{Dy}$&29.5101&28.8028&30.3917&27.9216&28.4526&28.4691&31.157 \\
&29.5086&28.6975&30.3727&27.9127&28.4691&28.5935& \\
$^{160}\textrm{W}\rightarrow ^{16}\textrm{O} +^{144}\textrm{Dy}$&30.1986&30.3007&31.0786&30.0405&29.5483&29.5663&31.927 \\
&30.2093&30.3126&31.0895&30.0739&29.5663&29.8251& \\
$^{162}\textrm{W}\rightarrow ^{16}\textrm{O} +^{146}\textrm{Dy}$&32.7798&32.5919&33.1311&32.8781&31.2475&31.2867&33.292 \\
&32.7960&32.5986&33.1432&32.8958&31.2867&32.4321& \\
$^{164}\textrm{W}\rightarrow ^{16}\textrm{O} +^{148}\textrm{Dy}$&37.2166&35.0472&37.9786&36.5910&33.7072&33.7293&34.362 \\
&37.2017&35.0540&37.9577&36.5940&33.7293&35.4355& \\
$^{166}\textrm{W}\rightarrow ^{16}\textrm{O} +^{150}\textrm{Dy}$&32.3325&32.3992&34.0840&32.2562&30.9283&30.9418&32.158 \\
&32.3346&32.4169&34.0689&32.2685&30.9418&31.8396& \\
$^{168}\textrm{W}\rightarrow ^{16}\textrm{O} +^{152}\textrm{Dy}$&29.7736&30.2490&31.4556&29.9717&29.1075&29.1339&29.962 \\
&29.7726&30.2534&31.4505&29.9762&29.1339&28.8249& \\
$^{170}\textrm{W}\rightarrow ^{16}\textrm{O} +^{154}\textrm{Dy}$&27.7213&28.7941&30.2402&28.8077&27.9051&27.1763&27.841 \\
&27.7212&28.8416&30.2304&28.7877&27.9333&-& \\
\\

$^{158}\textrm{W}\rightarrow ^{20}\textrm{Ne} +^{138}\textrm{Gd}$&37.4084
&37.8199&40.2825&37.1795&37.5530&37.0624&35.7945 \\
&37.4142&37.7133&40.2517&37.1381&37.5530&37.0619& \\
$^{160}\textrm{W}\rightarrow ^{20}\textrm{Ne} +^{140}\textrm{Gd}$&38.5361&37.8979&40.0555&37.2569&37.2671&37.2925&34.6913 \\
&38.5313&37.9152&40.0393&37.2690&37.2671&37.2945& \\
$^{162}\textrm{W}\rightarrow ^{20}\textrm{Ne} +^{142}\textrm{Gd}$&38.6017&38.6360&39.9717&38.4065&37.5541&37.8696&33.3599 \\
&38.6242&38.6508&39.9862&38.4521&37.5541&37.8534& \\
$^{164}\textrm{W}\rightarrow ^{20}\textrm{Ne} +^{144}\textrm{Gd}$&40.9146&40.2224&41.6357&40.6858&38.5794&39.8443&32.0253 \\
&40.9272&40.2396&41.6361&40.7305&38.5794&39.8258& \\
$^{166}\textrm{W}\rightarrow ^{20}\textrm{Ne} +^{146}\textrm{Gd}$&44.7203&42.0287&45.6637&43.7478&40.3151&-&30.5302 \\
&44.7027&42.0382&45.6402&43.7349&40.3151&-& \\
$^{168}\textrm{W}\rightarrow ^{20}\textrm{Ne} +^{148}\textrm{Gd}$&38.6716&38.8529&40.8899&38.7856&37.0623&38.0478&36.3782 \\
&38.6858&38.8553&40.8805&38.7894&37.0623&38.0478& \\

\hline
\end{tabular}}
\end{table}
\section{Results and discussion}
In the present study, we have made an attempt to study the feasibility of alpha and cluster decays from Tungsten (W) isotopes.  We have analysed all the possible parent-cluster combinations in tungsten isotopes.  From a detailed survey carried throughout the isotopic chain, it was found that only those nuclei belonging to the region between proton drip line and beta stability line exhibit these decay modes.


\subsection{Alpha decay}
At first, we have analysed the feasibility of alpha decay in this isotopic chain because of the availability of the wide range of experimental data.  It is found that W isotopes within the mass range 158 to 180 are unstable against alpha decay.  Any decay mode will be energetically favourable, if and only if the Q-value is positive.  $Q_{\alpha }$-values are calculated from binding energies using the relation,
\begin{equation}
Q_{\alpha} (N,Z)=B(N-2,Z-2)+B(2,2)-B(N,Z)
\end{equation}
where, B(N,Z) and B(N-2,Z-2) are the binding energies of the parent and the daughter nucleus ($_{72}\textrm{Hf}$).  B(2,2), the binding energy of $_{2}^{4}\textrm{He}$ nucleus (28.296 MeV) is taken from AME 2012 \cite{ame}.

\begin{figure}[h!]
\centerline{\includegraphics[width=3.0in]{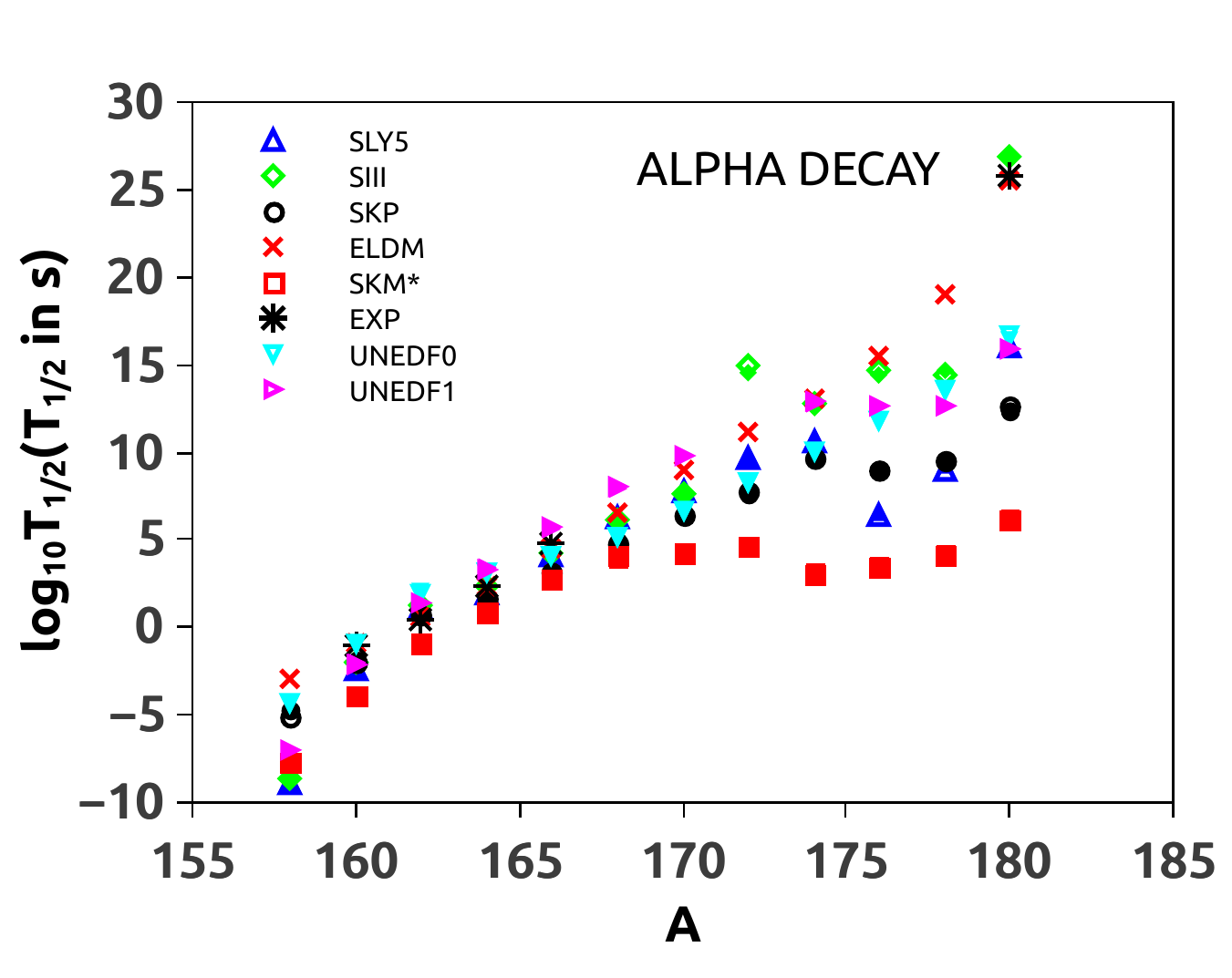}}
\vspace*{8pt}
\caption{Plots showing logarithmic value of half-life $(T_{1/2}$ in sec) against mass number of parent (A) nuclei corresponding to different decay modes for HO(solid) and THO(open) basis\protect\label{alph}.}
\end{figure}

The Q-values obtained in the case of different Skyrme forces are given in Table \ref{tab:alp}.  They have been compared with phenomenological Effective Liquid Drop Model (ELDM) \cite{eldm,pres} values as well as with the available experimental values\cite{exp,exp1}.  A small discrepency is observed in the estimated Q-values.  This is due to the fact that each Skyrme force describes binding energy of W isotopes with slight variation.  A small variation in the values of the parameters  of the Skyrme forces will affect the values of binding energy.  From Table \ref{tab:alp}, we can see that the values obtained by the recent parametrization, UNEDF0 and UNEDF1 as well as the classical Skyrme parametrization SIII agree with ELDM values.  Alpha decay half-lives are calculated using UDL.  Logarithmic value of half-lives against mass number of the parent(A) is depicted in Fig. \ref{alph}. From this figure, we can observe that the half-life is minimum for $^{158}\textrm{W}$, which leads to the magic daughter nuclei $^{154}\textrm{Hf}$ (N=82).  It is also visible that except SKM*, all other Skyrme forces overestimate the alpha decay half-lives.
 We have also studied the standard deviation of the half-lives with respect to experimental values for analysing how much the theoretical values agree with experimental ones.  Standard deviations are tabulated in Table \ref{tab:sd}.  From Table \ref{tab:sd}, it is observed that, SKM* is showing much variation with respect to experimental half-lives, compared to other Skyrme forces.

\begin{figure}[t]
\centerline{\includegraphics[width=5.0in]{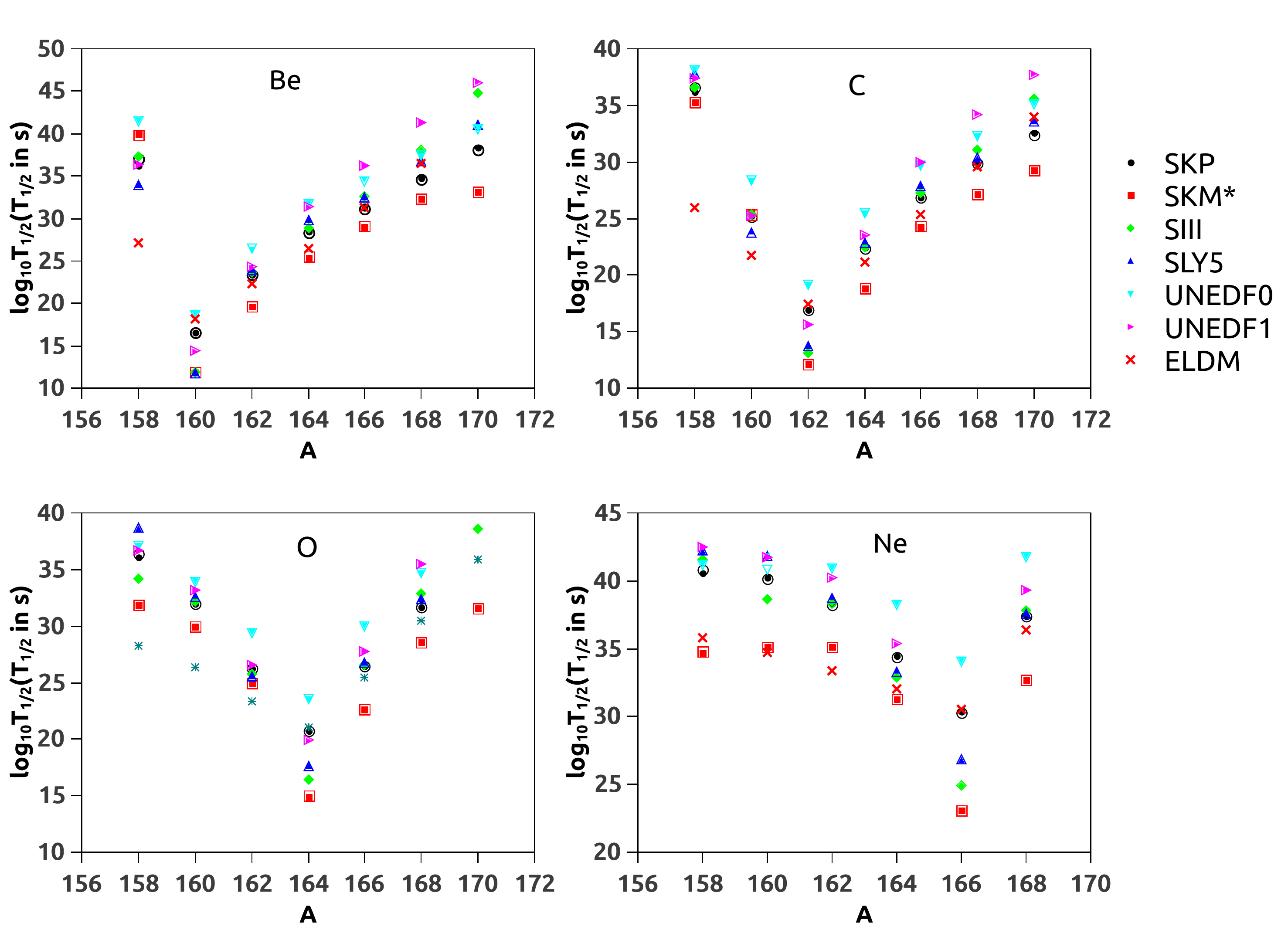}}
\vspace*{8pt}
\caption{Plots showing logarithmic value of half-life $(T_{1/2}$ in sec) against mass number of parent (A) nuclei, corresponding to different cluster decay modes for HO(solid) and THO(open) basis.\protect\label{clus}}
\end{figure}
\subsection{Cluster decay}
In W isotopes, we investigated the feasibility of the emission of different clusters like  $^{8}\textrm{Be}$,  $^{12}\textrm{C}$,  $^{16}\textrm{O}$ and $^{20}\textrm{Ne}$ .  The respective Q-values are estimated from binding energy using the following expressions,\\
$^{8}\textrm{Be}$:
\begin{equation}
Q(N,Z)=B(N-4,Z-4)+B(4,4)-B(N,Z)
\end{equation}
$^{12}\textrm{C}$:
\begin{equation}
Q(N,Z)=B(N-6,Z-6)+B(6,6)-B(N,Z)
\end{equation}
$^{16}\textrm{O}$:
\begin{equation}
Q(N,Z)=B(N-8,Z-8)+B(8,8)-B(N,Z)
\end{equation}
$^{20}\textrm{Ne}$:
\begin{equation}
Q(N,Z)=B(N-10,Z-10)+B(10,10)-B(N,Z)
\end{equation}
where, B(N-4,Z-4), B(N-6,Z-6), B(N-8,Z-8), B(N-10,Z-10) are the binding energies of the corresponding daughter nuclei ( $_{70}\textrm{Yb}$, $_{68}\textrm{Er}$, $_{66}\textrm{Dy}$ and $_{64}\textrm{Gd}$) and B(4,4), B(6,6), B(8,8) and B(10,10) are the binding energies of the clusters $^{8}\textrm{Be}$, $^{12}\textrm{C}$, $^{16}\textrm{O}$ and $^{20}\textrm{Ne}$ respectively.  Q-values calculated with respect to different Skyrme forces are given in Table \ref{tab:clus}.  We have compared the obtained results with the ELDM values.  In this case also we have calculated half-lives using UDL and they are depicted in Fig. \ref{clus}.

The decay rate for a particular decay mode will be maximum, if the corresponding half-life is minimum.  From Fig. \ref{clus}, it is found that in the case of $^{8}\textrm{Be}$ decay, the half-life is minimum for $^{160}\textrm{W}$.  This shows that the decay rate of $^{8}\textrm{Be}$ is maximum for $^{160}\textrm{W}$ isotope.  Also, this particular decay leads to the formation of the daughter nucleus, $^{152}\textrm{Yb}$ which is having magic neutron number(N=82).  Similarly for $^{12}\textrm{C}$, $^{16}\textrm{O}$, $^{20}\textrm{Ne}$ decay modes, half lives are minimum for those decays which leads to the formation of daughter nuclei (i.e, $^{150}\textrm{Er}$, $^{148}\textrm{Dy}$ and $^{146}\textrm{Gd}$) having magic neutron number (N=82).  These results show that the rate of decay will be maximum for those decay modes leading to magic daughter nuclei (N=82). These observations confirm the role of magicity in cluster decay. 

We have shown the values of half-lives predicted using different Skyrme forces in Fig. \ref{clus}.  All the calculations show similar trend in predicting the values, but with minor discrepancy in their magnitudes.  We have estimated the half-lives of all the decay modes from the binding energies of W isotopes, which are obtained using different Skyrme forces.  Each Skyrme force predicts the binding energy with a slight variation in its values and it is reflected in the predicted half-lives.  All the half-lives except those obtained by SKM* overestimate the ELDM values.  From Fig. \ref{clus}, it is observed that Ne radioactivity half-lives do not fall within the experimentally measurable range.  But we have predicted $^{20}\textrm{Ne}$ as well as $^{24}\textrm{Mg}$ decay in Os isotopes in our previous work\cite{os}and also in Pt isotopes.  This shows that as the parent nuclei becomes massive, we can expect more heavier cluster emissions. 


\begin{figure}[t]
\centerline{\includegraphics[width=5.5in]{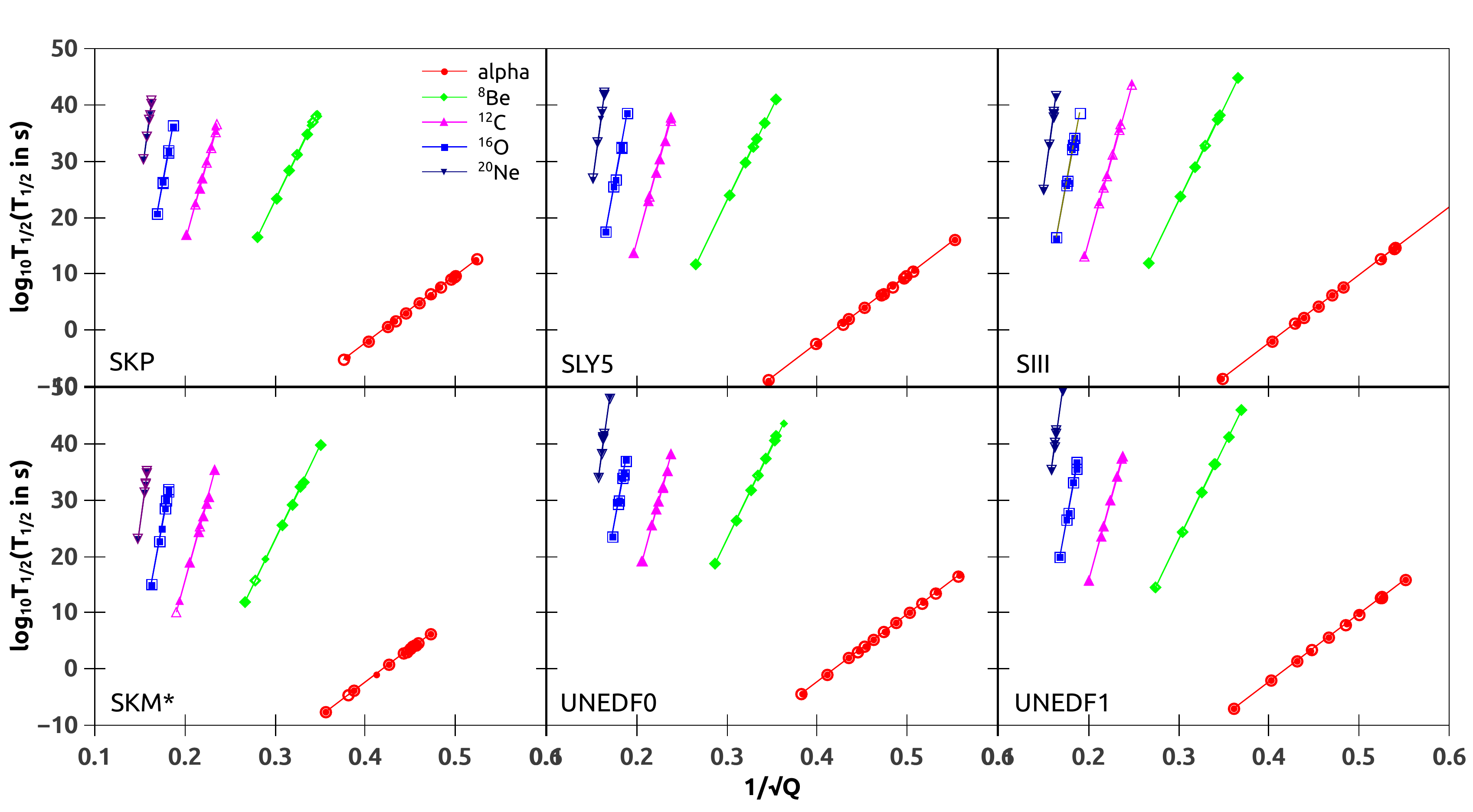}}
\vspace*{8pt}
\caption{Geiger-Nuttal plots of different cluster decay modes for HO(solid) and THO(open) basis corresponding to different Skyrme forces.\protect\label{gnp}}
\end{figure}

Geiger-Nuttel (GN) plot shows the relation between logarithmic half-lives and the disintegration energy(Q) of different decay modes.  Geiger-Nuttel law, which is a linear relation between these two quantities, is given by,
\begin{equation}
log_{10}T_{1/2}=\frac{X}{\sqrt{Q}}+Y
\end{equation}
where X and Y are the slopes and intercepts of the straight lines respectively.  Fig. \ref{gnp} shows the GN plots for different clusters emitted from W isotopes corresponding to various Skyrme parameters.  The linear nature of the plot is reproduced in the case of all the cluster modes.  Each emitted cluster has a specific slope and intercept.  They are given in Table \ref{tab:slope}.  From this table, we can see that as the emitted cluster becomes massive, the slope as well as the intercept increases.

\begin{table*}[t]
\caption{\label{tab:slope}Slopes and intercepts of even-even W isotopes calculated for different Skyrme forces }
\centering
\resizebox{\textwidth}{!}{ 
   \begin{tabular}{@{}*{13}{ccc}@{}} \hline
 \multicolumn{1}{c}{Skyrme}& \multicolumn{2}{c}{Alpha}& \multicolumn{2}{c}{Be}&\multicolumn{2}{c}{C}&\multicolumn{2}{c}{O}&\multicolumn{2}{c}{Ne}\\
force & Slope & Intercept & Slope & Intercept & Slope & Intercept & Slope & Intercept & Slope & Intercept \\ \hline
SKP & 119.866 & -50.212 & 330.189 & -76.060 & 580.124 & -100.322 & 879.117 & -127.861 & 1206.770 & -155.868\\
    & 119.803 & -50.182 & 330.406 & -76.060 & 580.572 & -100.422 & 879.117 & -127.861 & 1206.770 & -155.870\\
SLY5& 120.559 & -50.524 & 330.200 & -76.062 & 581.055 & -100.525 & 875.289 & -127.181 & 1190.031 & -153.194\\
    & 120.546 & -50.187 & 330.198 & -76.061 & 580.972 & -100.507 & 875.271 & -127.177 & 1193.551 & -153.683\\
SIII& 121.191 & -50.779 & 330.727 & -76.223 & 580.500 & -100.390 & 872.011 & -126.587 & 1184.827 & -152.336\\
    & 121.182 & -50.776 & 330.715 & -76.219 & 580.449 & -100.378 & 860.188 & -124.536 & 1185.006 & -152.364\\
SKM*& 119.353 & -50.063 & 331.083 & -76.390 & 583.610 & -101.112 & 866.964 & -125.747 & 1193.181 & -153.544\\
    & 119.605 & -50.179 & 331.793 & -76.589 & 583.614 & -101.114 & 865.863 & -125.553 & 1195.265 & -153.856\\
UNEDF0&120.241& -50.352 & 329.901 & -75.953 & 584.205 & -101.176 & 876.459 & -127.421 & 1135.247 & -144.43\\
    & 120.218 & -50.343 & 330.861 & -76.254 & 584.219 & -101.179 & 876.473 & -127.422 & 1132.531 & -143.99\\
UNEDF1&120.561& -50.492 & 330.377 & -76.093 & 582.262 & -100.744 & 871.549 & -126.519 & 1131.420 & -143.772\\
    & 120.545 & -50.485 & 330.357 & -76.087 & 582.248 & -100.741 & 871.161 & -126.531 & 1131.304 & -143.753\\ \hline

   \end{tabular} }
\end{table*}
 
\section{Conclusion}
In the present work, we have made an attempt to study the sensitivity of different Skyrme parametrizations in predicting the feasibility of alpha decay and cluster decay from W isotopes.  A comparitive study of half-lives was also done by using two different basis, HO and THO, which are used for solving HFB equations.  Both HO and THO basis do not produce much significant difference in the obtained results.  We have also compared the results with ELDM, which is a phenomenological model, wherever the experimental values are not available. 
 
We have observed that the trend of the half-lives predicted by different Skyrme forces are similar.  Depending on the various factors used for designing different Skyrme forces, the values of half-lives for different decay modes vary slightly.  We can get a qualitative description of alpha and cluster radioactivity using Skyrme HFB approach.  The most probable decay in various decay modes leads to magic daughter nuclei with N=82.  This again confirms the role of magicity in cluster radioactivity.  In the present study, we have selected only a few Skyrme forces.  The study can be extended to other Skyrme forces also.  It is hoped that these results can be a helping guide for the experimentalists in their future research.
\section*{Acknowledgments}

One of the authors, (NA) gratefully acknowledges UGC, Govt. of India, for providing the grant under UGC-JRF/SRF scheme and also for providing facilities made available under UGC-SAP-DRS II project.

\end{document}